# ENERGY DEPOSITION STUDIES FOR THE HI-LUMI LHC INNER TRIPLET MAGNETS [*][†]

N.V. Mokhov[#], I.L. Rakhno, S.I. Striganov, I.S. Tropin
*FNAL, Batavia IL 60510, USA*

F. Cerutti, L. Esposito, A. Lechner
*CERN, Geneva, Switzerland*

## Abstract

A detailed model of the High Luminosity LHC inner triplet region with new large-aperture $Nb_3Sn$ magnets, field maps, corrector packages, and segmented tungsten inner absorbers was built and implemented into the FLUKA and MARS15 codes. In the optimized configuration, the peak power density averaged over the magnet inner cable width is safely below the quench limit. For the integrated luminosity of 3000 $fb^{-1}$, the peak dose in the innermost magnet insulator ranges from 20 to 35 MGy. Dynamic heat loads to the triplet magnet cold mass are calculated to evaluate the cryogenic capability. In general, FLUKA and MARS results are in a very good agreement.

[*]Work supported by Fermi Research Alliance, LLC under contract No. DE-AC02-07CH11359 with the U.S. Department of Energy through the US LARP Program, and by the High-Luminosity LHC Project.
[†]Presented paper at the 5th International Particle Accelerator Conference, June 15-20, 2014, Dresden, Germany
[#]mokhov@fnal.gov

# ENERGY DEPOSITION STUDIES FOR THE HI-LUMI LHC INNER TRIPLET MAGNETS*

N.V. Mokhov#, I.L. Rakhno, S.I. Striganov, I.S. Tropin, FNAL, Batavia, IL 60510, USA
F. Cerutti, L. Esposito, A. Lechner, CERN, Geneva, Switzerland

## Abstract

A detailed model of the High Luminosity LHC inner triplet region with new large-aperture $Nb_3Sn$ magnets, field maps, corrector packages, and segmented tungsten inner absorbers was built and implemented into the FLUKA and MARS15 codes. In the optimized configuration, the peak power density averaged over the magnet inner cable width is safely below the quench limit. For the integrated luminosity of 3000 $fb^{-1}$, the peak dose in the innermost magnet insulator ranges from 20 to 35 MGy. Dynamic heat loads to the triplet magnet cold mass are calculated to evaluate the cryogenic capability. In general, FLUKA and MARS results are in a very good agreement.

## INTRODUCTION

After operation at the nominal luminosity, the LHC is planned to be upgraded to a 5-fold increased luminosity of $5\times10^{34}$ $cm^{-2}s^{-1}$. The upgrade includes replacement of the IP1/IP5 inner triplet (IT) 70-mm NbTi quadrupoles with the 150-mm coil aperture $Nb_3Sn$ quadrupoles along with the new 150-mm coil aperture NbTi dipole magnet.

As the first studies of radiation loads in the LHC upgrades have shown [1,2], one could provide the operational stability and adequate lifetime of the IT superconducting magnets by using the tungsten-based inner absorbers in the magnets. The goals are: (1) reduce the peak power density in the inner $Nb_3Sn$ cable to below the quench limit with a safety margin; (2) keep the 3000 $fb^{-1}$ lifetime peak dose in the innermost layers of insulation and radiation loads on inorganic materials in the hottest spots of the coils below the known radiation damage limits; (3) keep the dynamic heat load to the cold mass at a manageable level.

## FLUKA – MARS MODELING

To design such a system in a consistent and confident way, the coherent investigations have been undertaken with two independent Monte-Carlo codes benchmarked in the TeV energy region and used in such applications: FLUKA at CERN [3,4] and MARS15(2014) at Fermilab [5-7]. The studies were done for 7+7 TeV pp-collisions at the luminosity of $5\times10^{34}$ $cm^{-2}s^{-1}$ with a 295 μrad half-angle vertical crossing in IP1 (which was found earlier to be the worst case) using DPMJET-III as the event generator.

___________________________________________
*Work supported by Fermi Research Alliance, LLC, under contract No. DE-AC02-07CH11359 with the U.S. Department of Energy through the US LARP Program, and by the High-Luminosity LHC Project.
#mokhov@fnal.gov

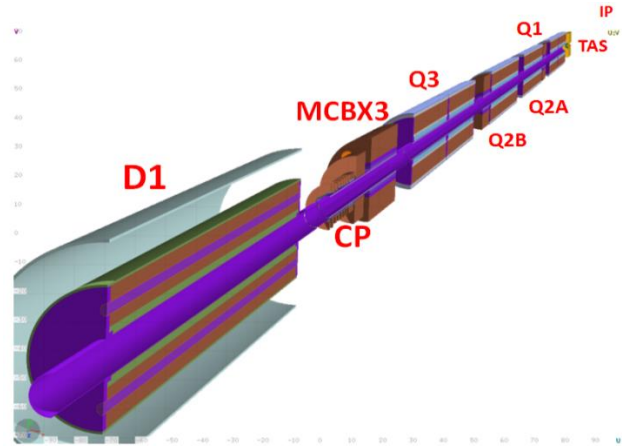

Figure 1: Computer model of HiLumi LHC inner triplet with correctors MCBX/CP and D1 dipole.

The identical, very detailed geometry model was created and used in both the codes with same materials and magnetic field distributions in each of the components of the 80-m region from IP through the D1 dipole. Figs. 1-3 show a 3D view of the model and details in the inner parts of the quadrupoles and orbit correctors. In this study, the maximum thickness of a segmented tungsten absorber in the first quadrupole is 1.6cm (up to ~32.5m from IP), and 0.6cm in the rest of IT and D1.

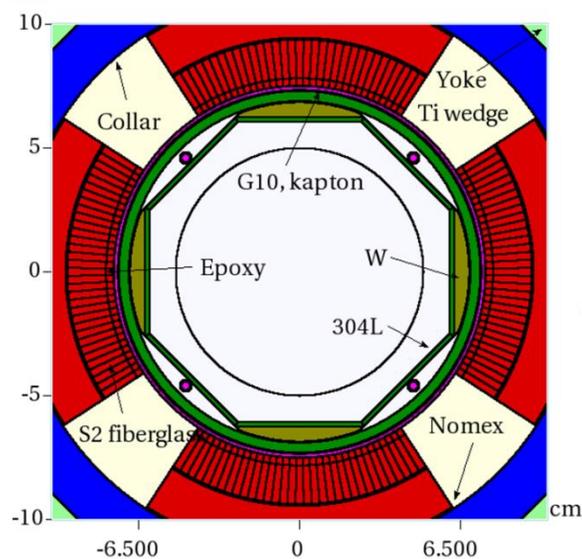

Figure 2: Details of the FLUKA-MARS model in the innermost region of the Q2-Q3 quadrupoles.

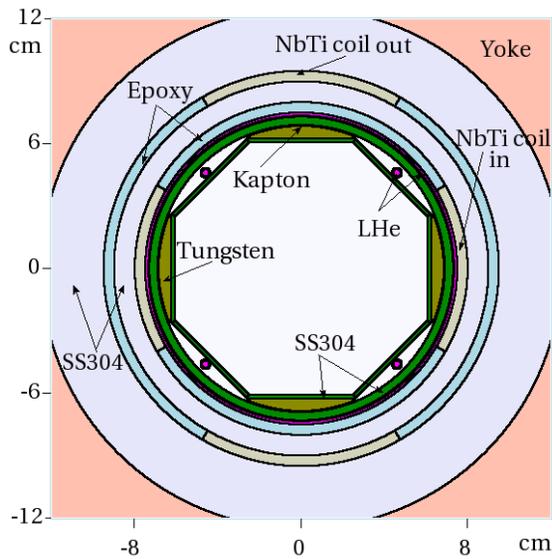

Figure 3: Cross-sectional view of the FLUKA-MARS model in the central part of the MCBX orbit correctors.

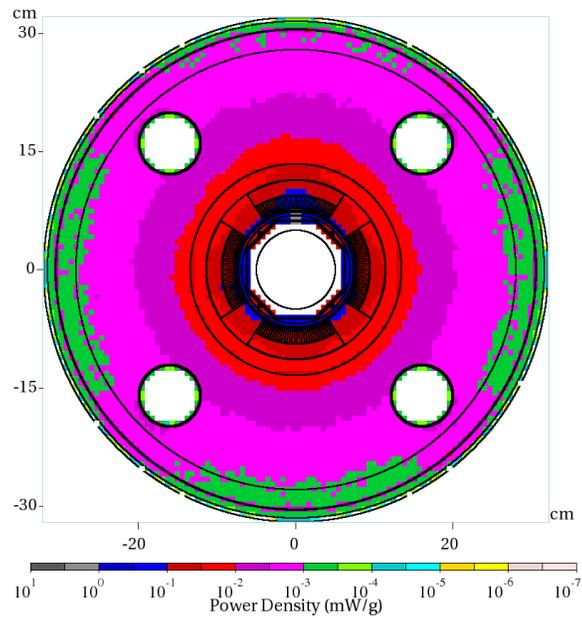

Figure 4: Power density isocontours at the IP end of Q2A.

Fine-mesh distributions of power density as well as of accumulated for 3000 fb$^{-1}$ integrated luminosity (~10-12 years of HiLumi LHC operation) absorbed dose, neutron fluence and Displacement-Per-Atom (DPA) along with dynamic heat load in every IT component were calculated with FLUKA and MARS in high-statistics runs. The power density and dynamic heat load results are normalized to the luminosity of $5\times10^{34}$ cm$^{-2}$s$^{-1}$, while all other ones to the 3000 fb$^{-1}$ integrated luminosity. Longitudinal scoring bins are 10 cm, and azimuthal ones are 2°. Radially, power density is scored in the superconducting cable width, while dose, fluence and DPA are scored at the azimuthal maxima within the innermost layer equal to 3-mm or its thickness, whatever is thinner.

## OPERATIONAL RADIATION LOADS

Power density isocontours at the IP end of the cold mass of the Q2A quadrupole are shown in Fig. 4. The longitudinal peak power density profile on the inner coils of the IT magnets at the azimuthal maxima is presented in Fig. 5. Results from FLUKA and MARS are in an excellent agreement. The peak value of 2 mW/cm$^3$ in the quadrupoles is 20 times less than the assumed quench limit of 40 mW/cm$^3$ in Nb$_3$Sn coils. The peak value of ~1.5 mW/cm$^3$ in the NbTi based coils of the correctors and D1 dipole is almost ten times less of the known quench limit 13 mW/cm$^3$ in such coils.

The total power dissipation in the IT region from the IP1 collision debris splits roughly 50-50 between the cold mass and the beam screen with tungsten absorber: 630 W and 615 W, respectively, from the FLUKA calculations. MARS predicts about 2% lower values. For the 45-m effective length of the cold mass, the average dynamic heat load on it is ~14 W/m.

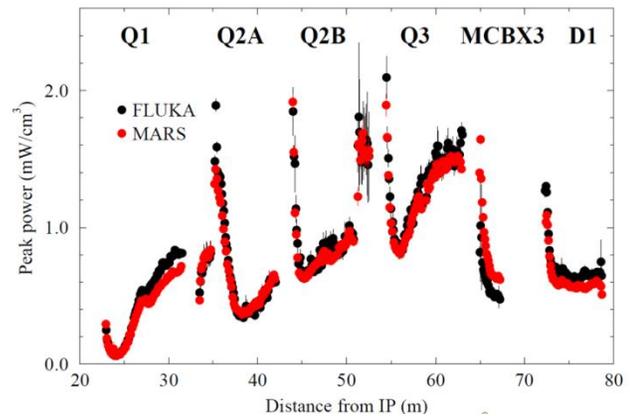

Figure 5: Longitudinal peak power density profile on the inner coils of the IT magnets.

## LIFETIME RADIATION LOADS

The peak dose and DPA – the quantities that define radiation damage and lifetime of insulators and non-organic materials of the IT magnets, respectively – are calculated at the azimuthal maxima in the innermost tiny layers of each the IT component shown in Figs. 2-3.

The longitudinal peak dose profiles on the inner coils and insulating materials are presented in Fig. 6. The values in the MCBX orbit correctors in the Q1-Q2A, Q2B-Q3 and Q3-D1 regions are given for the epoxy layer (FLUKA) and kapton layer (MARS); see Fig. 3 for details. Results from FLUKA and MARS are again in a good agreement. The larger aperture IT magnets and the tungsten absorbers implemented perform very well, reducing the peak values of both power density and absorbed dose in the HiLumi LHC IT to the levels which correspond to the LHC nominal luminosity.

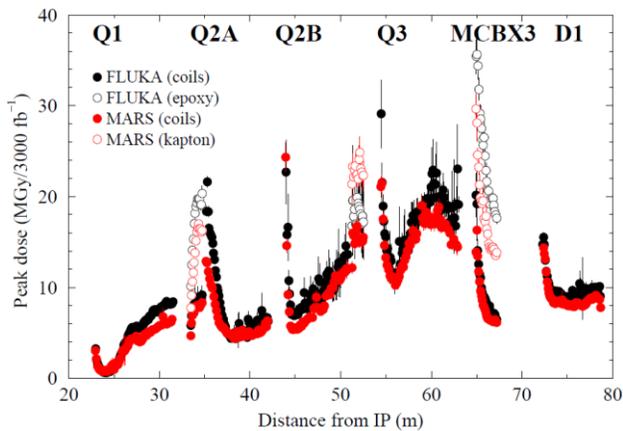 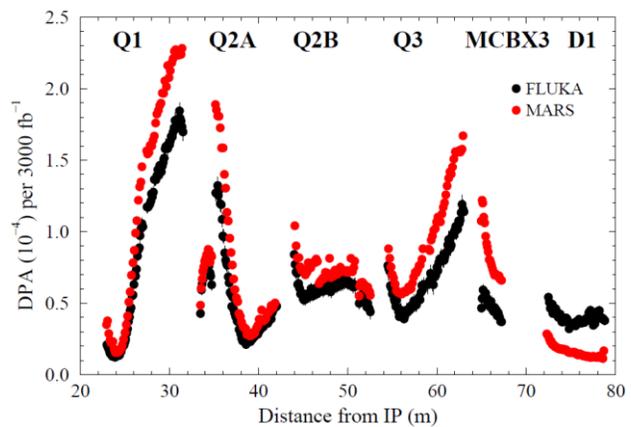

Figure 6: Longitudinal peak dose profile on inner coils and nearby insulators.

Figure 7: Longitudinal peak DPA profile along the hottest regions in the IT magnet coils.

The integrated peak dose in the IT magnet insulation reaches 30-36 MGy in the MCBX3 corrector, 28-30 MGy in the quadrupoles and ~22 MGy in the D1 dipole. This is at or slightly above the common limits for kapton (25-35 MGy) and CTD-101K epoxy (25 MGy). The maximum peak dose in the coils is about 25 MGy for quadrupoles and ~15 MGy for the D1 dipole.

Degradation of the critical properties of inorganic materials of the IT magnets – Nb3Sn and NbTi superconductors, copper stabilizer and mechanical structures – is usually characterized not by absorbed dose but by integrated neutron fluence and by DPA accumulated in the hottest spots over the expected magnet lifetime. DPA is the most universal way to characterize the impact of irradiation on inorganic materials. In both FLUKA and MARS, all products of elastic and inelastic nuclear interactions as well as Coulomb elastic scattering (NIEL) of transported charged particles (hadrons, electrons, muons and heavy ions) from ~1 keV to TeV energies contribute to DPA using Lindhard partition function and energy-dependent displacement efficiency. For neutrons at <20 MeV (FLUKA) and <150 MeV (MARS), the ENDF-VII database with NJOY99 processing is used in both the codes.

The longitudinal peak DPA profiles on the IT magnet coils are presented in Fig. 7. The peaks are generally observed at the inner coils; therefore, the data is given there. With the vertical crossing in IP1, the MCBX3 orbit corrector is the exception with the peak in the outer coil in the vertical plane (see Fig. 3). To see this effect, the MARS data in Fig. 7 for MCBX3 is given for the outer coil, while FLUKA shows results for the inner coil as in all other magnets. Contrary to the power density and dose distributions driven by electromagnetic showers initiated by photons from $\pi^0$ decays, DPA peaks at the non-IP end of the Q1B quadrupole. At that location, about 70% of DPA is from neutrons with E < 20 MeV, ~25% from transported nuclear recoils above 0.25 keV/A, and the rest is due to other transported particles and non-transported recoils.

The peak in the Q1B inner coil is about $2\times10^{-4}$ DPA per 3000 fb$^{-1}$ integrated luminosity. In other IT components it is about $(1\pm0.5)\times10^{-4}$. These numbers should be acceptable for the superconductors and copper stabilizer provided periodic annealing at the collider shutdowns. Taking into account a good correlation of DPA with neutron fluence in the coils, one can also compare the latter with the known limits. In the quadrupole coils, the peak fluence is $\sim2\times10^{17}$ cm$^{-2}$ which is substantially lower than the $3\times10^{18}$ cm$^{-2}$ limit used for the Nb$_3$Sn superconductor. In the orbit corrector and D1 dipole coils, the peak fluence is $\sim5\times10^{16}$ cm$^{-2}$ which is again lower than the $10^{18}$ cm$^{-2}$ limit used for the NbTi superconductor.

The integrated DPA in the magnet mechanical structures are 0.003 to 0.01 in the steel beam screen and tungsten absorber, $\sim 10^{-4}$ in the collar and yoke, and noticeably less outside. These are to be compared to a ~10 DPA limit for mechanical properties of these materials. Neutron fluences in the IT mechanical structures range from $3\times10^{16}$ cm$^{-2}$ to $3\times10^{17}$ cm$^{-2}$ compared to the $10^{21}$ cm$^{-2}$ to $7\times10^{22}$ cm$^{-2}$ limits.